\documentclass[]{spie}  

\usepackage{amsmath,amsfonts,amssymb}
\usepackage{graphicx}
\usepackage[colorlinks=true, allcolors=blue]{hyperref}
\usepackage{multirow}
\usepackage{array}
\usepackage{longtable}
\usepackage{soul}
\usepackage{enumitem}

\usepackage{xcolor,colortbl}
\usepackage[]{aas_macros}

\graphicspath{{figures/}}

\title{The final design of GMagAO-X: the wavefront sensing and control (WFS\&C) architecture of GMagAO-X}

\author{
Sebastiaan Y. Haffert$^{*,a,b}$,
Jared R. Males$^{b}$,
Laird M. Close$^{b}$,
Olivier Guyon$^{b,c}$,
Olivier Durney$^{b}$,
Maggie Kautz$^{b}$,
Liam Koning$^{a}$,
Louis Desdoigts$^{a}$,
Matthijs Mars$^{a}$,
Elena Tonucci$^{a}$,
Adam K. Taras$^{a}$,
Yinzi Xin$^{a}$,
Rico Landman$^{a}$
\\
\vspace{0.2cm}
$^{a}$Leiden Observatory, Leiden University, The Netherlands\\
$^{b}$Steward Observatory, University of Arizona, USA\\
$^{c}$Subaru Telescope / National Astronomical Observatory of Japan
}

\authorinfo{*sebastiaan.haffert@strw.leidenuniv.nl}

\begin{document} 
\maketitle


\begin{abstract}The Giant Magellan Adaptive Optics eXtreme (GMagAO-X) instrument has now been selected as an official part of the Giant Magellan Telescope’s (GMT) instrument suite. The instrument will be ready at first-light of the GMT in the mid 2030s. The instrument is now progressing towards its final design with a final design review planned for March, 2027. The high-density actuator deformable mirror with 21.000 actuators will allow GMagAO-X to create diffraction-limited images from visible to near-infrared. GMagAO-X will be coupled with high-performance coronagraphs to search for exoplanets at the diffraction-limit. The coronagraphs require wavefront control at sub-nm precision. We will provide an update on the wavefront sensing and control architecture and how the loops interact with each other through end-to-end simulations.\end{abstract}



\section{Introduction}

The next generation of Extremely Large Telescopes (ELTs) will enable direct imaging and spectroscopic characterization of rocky exoplanets around nearby stars \cite{kasper2021pcs,males2024gmagaox}. In particular, the 25.4\,m Giant Magellan Telescope (GMT) will provide sufficient angular resolution to observe the habitable zones of the nearest M-dwarf systems at visible wavelengths\cite{males2024gmagaox}. Achieving this science case requires diffraction-limited imaging at separations of only a few $\lambda/D$ together with raw contrasts approaching $10^{-5}$, enabling post-processing to reach final contrasts of $\sim10^{-8}$.

GMagAO-X has been developed specifically to address this science case\cite{males2024gmagaox,close_2024,haffert_2024b}. Building on the successful operation of MagAO-X\cite{males_2024}, the instrument combines high-order adaptive optics, advanced coronagraphy, focal-plane wavefront sensing and telemetry-driven control into a single integrated system optimized for visible-light high-contrast imaging. Since the Preliminary Design Review in 2024, the instrument has matured significantly. The optical and opto-mechanical design has converged to its Final Design Review (FDR) configuration, while the wavefront sensing and control (WFS\&C) architecture has evolved into a hierarchy of coupled feedback loops that actively stabilize every critical optical degree of freedom.

Unlike classical adaptive optics systems, GMagAO-X does not rely on a single control loop. Instead, independent control loops operate on different spatial and temporal scales. The visible or infrared Pyramid Wavefront Sensor (PyWFS) provides high-order atmospheric correction, the Holographic Dispersed Fringe Sensor (HDFS) continuously measures differential piston between the GMT segments, predictive controllers suppress structural vibrations, focal-plane wavefront sensors maintain the coronagraphic dark hole, and telemetry from all sensors is archived to enable advanced post-processing and controller optimization.

The philosophy underlying the instrument is straightforward:

\begin{quote}
\emph{Every optical element that requires tight stability is actively controlled.}
\end{quote}

This design philosophy minimizes quasi-static aberrations, reduces calibration overhead, and naturally lends itself to a hierarchical control architecture in which each sensor operates on the optical aberrations for which it provides the highest sensitivity.

This paper presents the current WFS\&C architecture of GMagAO-X. Rather than describing every subsystem in detail, we focus on the interaction between the various feedback loops and their role in delivering diffraction-limited performance. The detailed opto-mechanical implementation of the instrument is presented separately by Close et al.\cite{Close2026GMagAOXFDR}, while the individual wavefront sensing technologies are described in dedicated companion papers.

\section{Overview of the Wavefront Sensing and Control Architecture}

\begin{figure}
\includegraphics[width=\textwidth]{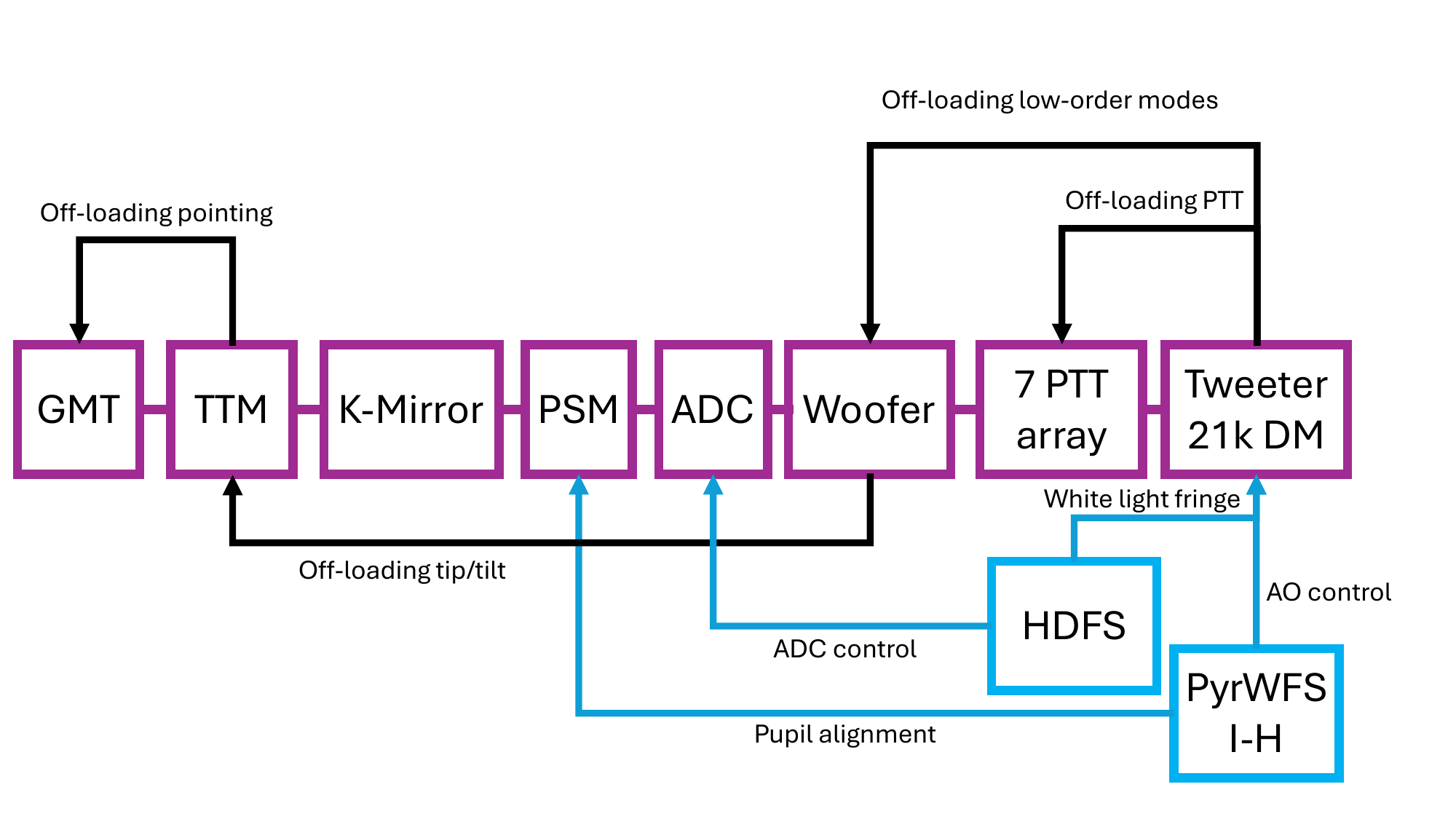}
\caption{The architecture of the high-order AO loop. The light from the GMT is first tip/tilt stabilized after which it is sent through the K-mirror. The K-mirror provides a stabilized pupil. A pupil steering mirror (PSM) is used to correct the pupil alignment for potential wobbles or drifts by the K-mirror. The two counter-rotating ADC prisms provide common path correction of the atmospheric dispersion. A woofer-tweeter architecture is then used to control the atmospheric aberrations. The GMagAO-X beam is split up into 7 paths each containing its own piston/tip/tilt piezo mirror and 3K BMC DM. The blue lines show the wavefront sensors that provide direct control feedback while the black lines indicate off-loading control.}
\label{fig:architecture}
\end{figure}

Figure~\ref{fig:architecture} illustrates the overall wavefront sensing and control architecture adopted for GMagAO-X. Incoming light from the GMT first passes through a fast tip--tilt mirror, K-mirror derotator and atmospheric dispersion compensator before reaching the woofer and 21\,000 actuator parallel deformable mirror\cite{close_2024}. Depending on the observing mode, visible or infrared light is directed toward a Pyramid Wavefront Sensor operating at kilohertz frame rates.

Approximately ten percent of the near-infrared light is simultaneously directed toward the Holographic Dispersed Fringe Sensor (HDFS)\cite{10.1117/1.JATIS.8.2.021513}, which continuously measures differential piston between the seven GMT primary mirror segments. Downstream of the adaptive optics system, science light enters the coronagraph where additional focal-plane wavefront sensors monitor non-common-path aberrations and maintain the deep coronagraphic null.

The architecture naturally separates the control problem according to spatial and temporal frequency. Atmospheric turbulence is corrected by the high-order adaptive optics loop, structural vibrations are mitigated through predictive control (Haffert et al. 2026 and Johnson et al. 2026, this proceeding), slow segment phasing errors are removed using the HDFS \cite{10.1117/1.JATIS.8.2.021513}, while non-common-path aberrations are corrected directly at the science focal plane. This modular approach reduces coupling between individual controllers while allowing information to be exchanged through a common telemetry framework.

A distinguishing feature of GMagAO-X is that all telemetry generated by these control loops is archived. This enables off-line performance analysis, telemetry-driven post-processing, and the future implementation of data-driven control algorithms that can exploit correlations between the various sensors.

\subsection{High-order adaptive optics}

The primary adaptive optics loop of GMagAO-X is built around an unmodulated Pyramid Wavefront Sensor (PyWFS) operating at kilohertz frame rates. While modulation has traditionally been employed to increase the linear capture range of the PyWFS, it comes at the expense of sensitivity by reducing the optical gain. As the wavefront error decreases, an unmodulated PyWFS provides significantly greater sensitivity to small aberrations, making it particularly well suited for the high-Strehl regime required for coronagraphic imaging.

Historically, the nonlinear response of an unmodulated PyWFS has limited its application because conventional reconstruction algorithms assume a linear relationship between the measured pupil intensities and the incoming wavefront. These assumptions break down once the residual wavefront exceeds a small fraction of a wavelength, requiring modulation or complex nonlinear optimization techniques.

Recent advances in machine learning provide an attractive alternative\cite{landman2024making,landman2025making}. Rather than explicitly linearizing the sensor response, convolutional neural networks (CNNs) can learn the nonlinear mapping between the four pyramid pupil images and the underlying wavefront directly from simulated and laboratory calibration data. The network effectively performs nonlinear wavefront reconstruction while maintaining the computational throughput required for real-time control\cite{landman2025making}.

\begin{figure}
\includegraphics[width=\textwidth]{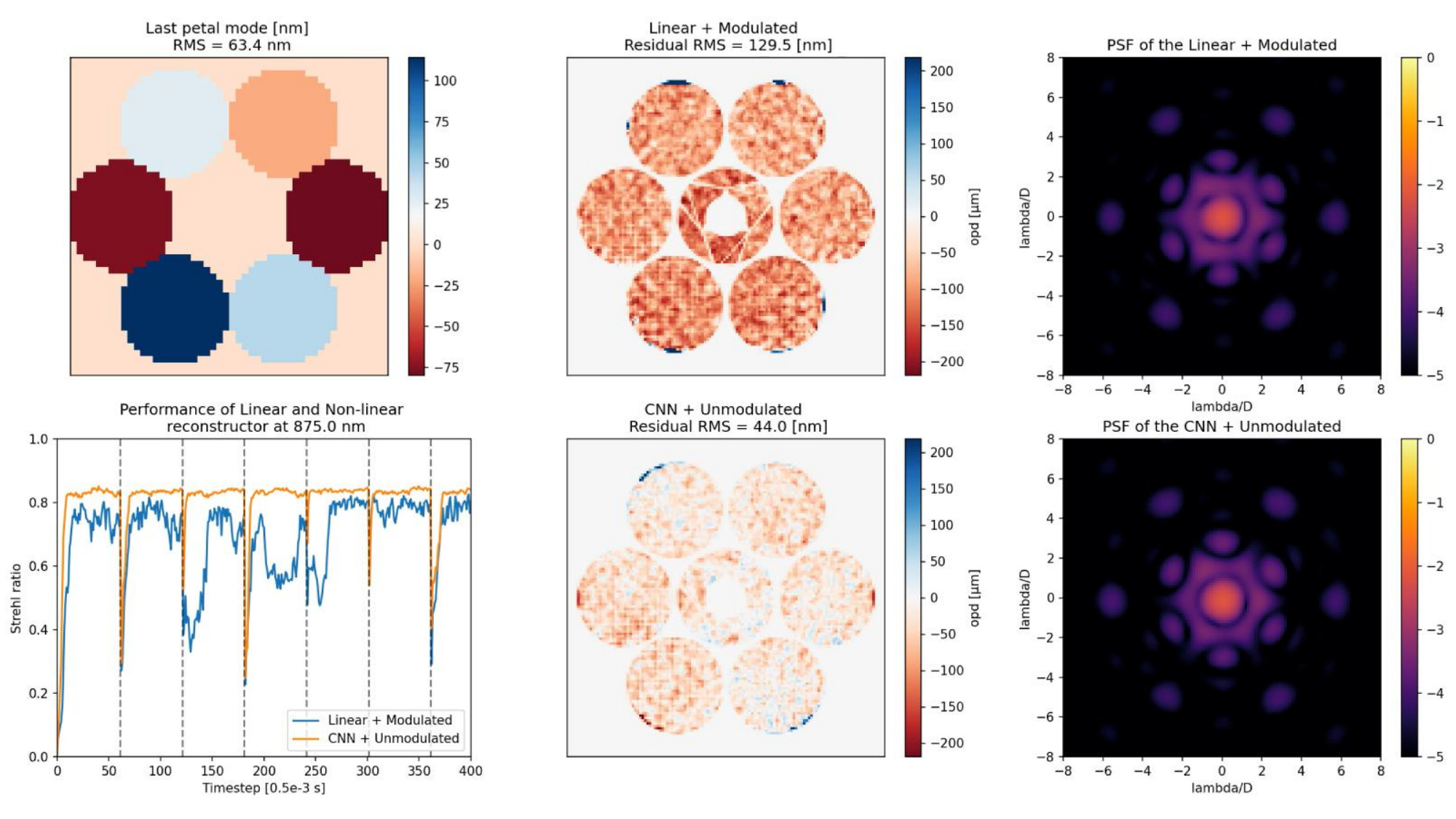}
\caption{Initial end-to-end simulations of a neural network for differential piston and atmospheric control for the GMT. Top left shows the latest injected differential piston signal. The bottom left shows the Strehl as a function of time. A new set of differential pistons is injected at each of the vertical dashed gray lines. The middle and right columns show the phase residuals and the corresponding PSF for the modulated pyramid on the top and the unmodulated pyramid on the bottom. The performance of the unmodulated pyramid wavefront sensor is significantly more stable than that of the modulate pyramid wavefront sensor. It maintaints a high Strehl ($>$0.8) and recovers quickly from segment phase errors.}
\label{fig:cnn}
\end{figure}

For GMagAO-X, the CNN reconstructor enables operation of the PyWFS without modulation while maintaining a sufficiently large capture range to acquire and stabilize atmospheric turbulence. In Figure \ref{fig:cnn}, we show the improved stability of the neural network performance in the presence of atmospheric residuals and differential piston. The increased sensitivity of the unmodulated sensor directly translates into improved wavefront estimation accuracy at high Strehl ratios, reducing residual wavefront error at the science focal plane.

An important consequence of the nonlinear reconstruction is that the PyWFS is no longer restricted to correcting atmospheric turbulence alone. The network is trained to estimate both conventional atmospheric aberrations and piston discontinuities between the seven GMT segments. Consequently, the high-order adaptive optics loop simultaneously corrects atmospheric turbulence and high-frequency differential piston using the 21\,000 actuator parallel deformable mirror. This significantly extends the temporal bandwidth of segment phasing beyond what is achievable using dedicated piston sensors alone.

The Holographic Dispersed Fringe Sensor (HDFS) therefore assumes a complementary role within the overall control architecture \cite{10.1117/1.JATIS.8.2.021513, kautz2024phasing}. Rather than operating as the primary piston controller, the HDFS provides robust measurements of the absolute segment piston over a large dynamic range, allowing the system to acquire and maintain phase coherence even when large piston offsets are present. Once the segmented pupil has been co-phased, the unmodulated PyWFS maintains fine piston correction at kilohertz bandwidth together with the atmospheric adaptive optics correction. The two sensors therefore operate over complementary temporal and dynamic ranges: the HDFS provides robust absolute phasing, while the PyWFS suppresses the fast residual piston errors that ultimately limit coronagraphic performance. This architecture was recently tested on-sky with the MagAO-X instrument \cite{kautz2024phasing}.

This hierarchical architecture combines the large capture range of the HDFS with the high sensitivity of the unmodulated PyWFS, enabling continuous diffraction-limited operation while minimizing the non-common-path errors between piston sensing and the science beam.


\subsection{Active atmospheric dispersion correction}

Residual atmospheric dispersion is a significant source of performance degradation for visible-light extreme adaptive optics systems. Even after applying the geometrical correction predicted by the atmospheric dispersion compensator (ADC) model, small residuals remain due to uncertainties in the atmospheric conditions, telescope pointing, prism calibration, and chromatic optical alignment. These residuals broaden the science point spread function, reduce the sensitivity of the Pyramid Wavefront Sensor, and ultimately limit coronagraphic contrast.

Rather than relying exclusively on an open-loop atmospheric model, GMagAO-X implements a closed-loop correction of the residual atmospheric dispersion (Twitchell et al. 2026, this proceedings). The Holographic Dispersed Fringe Sensor (HDFS), originally developed for segment piston sensing, naturally provides a measurement of the chromatic displacement of the dispersed interference fringes. Since the expected fringe pattern is accurately known, residual atmospheric dispersion appears as a wavelength-dependent displacement of the fringes relative to their calibrated positions (Desdoigts et al in prep.).

This chromatic offset is estimated continuously during science observations and converted into low-order correction commands for the ADC. Because the HDFS operates simultaneously with the high-order adaptive optics loop, the residual atmospheric dispersion can be monitored without interrupting science observations or introducing additional calibration exposures.

The resulting control architecture separates the atmospheric dispersion correction into two components. An open-loop model computes the bulk prism rotation based on telescope elevation and atmospheric parameters, while the HDFS provides a slow feedback signal that removes residual chromatic errors. This hybrid approach combines the large capture range of the geometric ADC model with the precision of focal-plane telemetry, ensuring optimal image quality under changing atmospheric conditions.


\subsection{Pupil steering and deformable mirror registration}

Maintaining accurate registration between the GMT pupil, the Pyramid Wavefront Sensor, and the deformable mirrors is essential for stable high-order wavefront correction. Small pupil shifts reduce reconstruction accuracy, introduce modal cross-talk, and degrade the performance of both atmospheric correction and segment phasing.

GMagAO-X therefore continuously monitors the pupil alignment using the Pyramid Wavefront Sensor itself. The pupil images produced by the PyWFS provide a direct measurement of the illuminated telescope pupil and allow slow translations and rotations of the pupil to be estimated with sub-pixel precision. These measurements are combined with the commanded actuator patterns on the deformable mirrors to estimate the relative registration between the reconstructed wavefront and the physical actuator geometry.

During normal operation, calibration patterns naturally present in the deformable mirror telemetry provide additional information about the pupil location. By comparing the expected response of these actuator commands with the measured PyWFS pupil images, the control system estimates the pupil offset with respect to the deformable mirror influence function model. This estimator operates continuously in the background without requiring dedicated calibration sequences.

The estimated pupil offset is removed by commanding the pupil steering mirrors located upstream of the wavefront sensor. Because the correction is applied optically rather than numerically, the reconstruction matrix and neural-network reconstructor remain valid throughout the observation. This strategy minimizes calibration overhead while maintaining optimal registration between the telescope pupil, the wavefront sensor, and the 21\,000 actuator parallel deformable mirror.

Together with the high-order adaptive optics loop and the HDFS phasing loop, the pupil steering controller forms another element of the hierarchical WFS\&C architecture. Each loop operates on a different degree of freedom---wavefront phase, segment piston, atmospheric dispersion, or pupil position---allowing the instrument to maintain a fully calibrated optical system throughout long science observations.

\section{Coronagraphic wavefront control}

The high-order adaptive optics system delivers a diffraction-limited wavefront to the coronagraph. However, achieving the raw contrasts required for exoplanet imaging requires substantially tighter wavefront stability than is needed to obtain a high Strehl ratio alone. Slowly evolving non-common-path aberrations, thermally induced drifts, mechanical flexure, and residual pointing errors all produce quasi-static speckles that cannot be distinguished from faint astrophysical companions.

Rather than relying on the high-order AO deformable mirrors for coronagraph optimization, GMagAO-X employs two dedicated high-order non-common-path deformable mirrors (NCP DMs), each with approximately 3000 actuators, located immediately upstream of the coronagraphic optics. These deformable mirrors are used exclusively for coronagraphic wavefront control and are therefore decoupled from the atmospheric adaptive optics system. This is a similar architecture that has now been succesfully demonstrated on-sky with MagAO-X\cite{kueny2024magao,haffert2026sky}. While the upstream woofer and 21\,000 actuator parallel deformable mirror correct atmospheric turbulence, telescope vibrations, and segment piston, the dedicated NCP DMs continuously compensate the slowly varying aberrations introduced by the downstream optics. This separation of responsibilities minimizes coupling between the atmospheric and coronagraphic control loops while allowing each controller to operate at the temporal bandwidth most appropriate for the aberrations it corrects.

The coronagraphic control architecture is naturally divided into two regimes. Low-order aberrations, such as pointing, focus, astigmatism, and slow pupil drifts, are corrected using dedicated focal-plane wavefront sensors operating continuously during science observations. Residual high-order aberrations are removed using implicit Electric Field Conjugation (iEFC) \cite{2023A&A...673A..28H}, which estimates and suppresses the coherent stellar electric field responsible for quasi-static speckles. Together these loops maintain the deep coronagraphic null required for high-contrast imaging over long observing sequences.


\subsection{Low-order wavefront sensing using FLOWFS and LLOWFS}

Residual low-order aberrations dominate the temporal stability of the coronagraphic point spread function (PSF). Even small pointing drifts or focus variations result in leakage of stellar light through the coronagraph, rapidly degrading the achievable contrast. Since these aberrations originate downstream of the main adaptive optics system, they cannot be measured by the Pyramid Wavefront Sensor and therefore require dedicated focal-plane sensing.

GMagAO-X adopts complementary focal-plane wavefront sensors for both the focal plane and Lyot plane. The Focal Plane Low-Order Wavefront Sensor (FLOWFS) measures residual aberrations directly from the rejected starlight at the coronagraphic focal plane mask\cite{Guyon2009}, providing high sensitivity to pointing, focus, and other low-order modes that affect coronagraph throughput. We have demonstrated this at 8 kHz with the MagAO-X instrument (Mars et al. in prep), which increased the stability of our coronagraphic alignment to the milli arcsecond level. The Lyot Low-Order Wavefront Sensor (LLOWFS) monitors the starlight rejected by the reflective Lyot stop, providing an estimate of residual wavefront errors for phase-shifting coronagraphs. Our prime choice coronagraph for small-inner working angles will be the Phase Induced Amplitude Apodization Complex Mask Coronagraph (PIAACMC) \cite{guyon2014high, tonucci2026phase}.

These sensors operate continuously during science observations and therefore provide a direct measurement of the aberrations that limit coronagraphic performance. Their measurements are converted into commands for the dedicated non-common-path deformable mirrors located immediately upstream of the coronagraph. Because these mirrors are downstream of the high-order adaptive optics system, they correct only the residual aberrations seen by the science beam without perturbing the upstream atmospheric correction. The resulting control loop actively stabilizes the coronagraphic PSF over long integrations, minimizing temporal drift of the stellar leakage and providing a stable starting point for high-order focal-plane wavefront control.

The measurements from FLOWFS and LLOWFS are converted into commands for the dedicated 3k actuator non-common-path deformable mirrors located immediately upstream of the coronagraph. Because these mirrors are reserved exclusively for coronagraphic wavefront control, they can continuously compensate slowly evolving optical drifts without perturbing the high-order atmospheric correction performed by the upstream adaptive optics system. This architecture effectively separates the atmospheric and coronagraphic control problems, allowing both systems to converge independently while sharing the same science beam.


\subsection{High-order focal-plane wavefront control using Electric Field Conjugation}

Once the low-order aberrations have been stabilized, the dominant limitation to raw contrast becomes coherent stellar speckles generated by residual high-order non-common-path aberrations. These speckles evolve slowly with time and remain coherent with the stellar electric field, making them amenable to focal-plane wavefront control.

GMagAO-X employs Electric Field Conjugation (EFC) to estimate the complex electric field within the science image and suppress coherent speckles inside the dark-hole region surrounding the star \cite{haffert2026sky}. Small probe patterns are introduced on the non-common-path deformable mirror to modulate the stellar electric field. The resulting intensity changes recorded by the science camera are used to estimate the complex amplitude of the coherent speckles, after which the deformable mirror is commanded to generate an equal-amplitude field with opposite phase.

The dedicated 3k actuator NCP DMs provide the spatial control authority required to sculpt high-contrast dark holes while preserving the atmospheric correction provided by the upstream adaptive optics system. Since these mirrors are not used for atmospheric correction, their full dynamic range and actuator authority can be devoted to minimizing coherent stellar leakage in the science focal plane. This decoupling substantially simplifies the control architecture and enables continuous optimization of the coronagraph without introducing disturbances into the high-bandwidth adaptive optics loop.

Together, the low-order focal-plane sensors and high-order EFC controller complete the hierarchical wavefront sensing and control architecture of GMagAO-X. The atmospheric adaptive optics system provides the diffraction-limited input wavefront, the HDFS maintains segment phasing, the ADC and pupil controllers preserve optical alignment, while the coronagraphic control loops remove the final residual aberrations that ultimately determine the achievable raw contrast.

\section{Conclusion and future work}

The wavefront sensing and control architecture of GMagAO-X has now converged to its final design in preparation for the instrument's Final Design Review in 2027. The architecture combines a hierarchy of complementary control loops operating over different spatial and temporal scales, including high-order atmospheric correction with an unmodulated Pyramid Wavefront Sensor, high-bandwidth segment piston correction, active atmospheric dispersion compensation, continuous pupil registration, and dedicated coronagraphic wavefront control using low-order focal-plane sensing and iterative Electric Field Conjugation. Together these systems provide a unified control framework capable of maintaining the wavefront stability required for visible-light high-contrast imaging on the Giant Magellan Telescope.

A major strength of the GMagAO-X architecture is that it is built upon technologies that have already been demonstrated individually on-sky with MagAO-X. The Pyramid Wavefront Sensor, predictive control, Holographic Dispersed Fringe Sensor, focal-plane wavefront sensing, and coronagraphic control algorithms have all been developed and validated on the Magellan Clay telescope under realistic observing conditions. Rather than introducing fundamentally new sensing techniques, GMagAO-X integrates these mature technologies into a unified control architecture tailored to the segmented aperture and increased actuator count of the Giant Magellan Telescope. This evolutionary approach substantially reduces technical risk while leveraging nearly a decade of operational experience with visible-light extreme adaptive optics.

The semi-analytical performance analysis presented in the 2024 GMagAO-X SPIE proceedings \cite{males2024gmagaox,haffert_2024b} established the expected performance of the proposed wavefront sensing and control architecture and guided many of the design decisions that have led to the final instrument configuration presented here. With the architecture now finalized, the next phase of the project is to verify these performance predictions using detailed end-to-end simulations incorporating realistic atmospheric turbulence, telescope vibrations, detector noise, optical propagation, control-loop interactions, and non-common-path aberrations. These simulations will quantify the performance of the fully integrated control architecture, validate the assumptions of the earlier analytical models, and provide the final optimization of the real-time control system before the instrument enters construction.

The convergence of both the opto-mechanical design and the wavefront sensing and control architecture marks an important milestone in the development of GMagAO-X. As the project progresses toward Final Design Review, the focus now shifts from architectural trade studies to quantitative verification of system performance. By combining an architecture built upon individually demonstrated on-sky technologies with comprehensive end-to-end performance validation, GMagAO-X is well positioned to deliver the wavefront stability required to exploit the full diffraction-limited capabilities of the Giant Magellan Telescope for high-contrast exoplanet science.

\acknowledgments 
The GMagAO-X conceptual and preliminary design would not have been possible without the support of the University of Arizona Space Institute.  We are also grateful for the support of an anonymous donor to Steward Observatory.

\bibliography{report} 
\bibliographystyle{spiebib} 

\end{document}